\documentclass[twocolumn,prl,showpacs,superscriptaddress]{revtex4}
\usepackage{amssymb}
\usepackage{amsmath}
\usepackage{graphicx}
\usepackage{subfigure}
\usepackage{natbib}
\usepackage{epsfig}
\usepackage{amsfonts}
\usepackage{mathrsfs}
\usepackage{CJK}
\usepackage[toc,page,title,titletoc,header]{appendix}

\begin{document}
\title{Quantum speed limit for arbitrary initial states}

\author{Yingjie Zhang }
 \email{qfyingjie@iphy.ac.cn}
 \affiliation{Institute of
Physics, Chinese Academy of Sciences, Beijing, 100190, China}
\affiliation{Shandong Provincial Key Laboratory of Laser
Polarization and Information Technology, Department of Physics, Qufu
Normal University, Qufu 273165, China}
\author{Wei Han }
 \affiliation{Shandong Provincial Key
Laboratory of Laser Polarization and Information Technology,
Department of Physics, Qufu Normal University, Qufu 273165, China}
\author{Yunjie Xia }
 \affiliation{Shandong Provincial Key
Laboratory of Laser Polarization and Information Technology,
Department of Physics, Qufu Normal University, Qufu 273165, China}
\author{Junpeng Cao }
 \affiliation{Institute of Physics,
Chinese Academy of Sciences, Beijing, 100190, China}
\author{Heng Fan }
\email{hfan@iphy.ac.cn}
 \affiliation{Institute of Physics,
Chinese Academy of Sciences, Beijing, 100190, China}

\date{\today}
\begin{abstract}
We investigate the generic bound on the minimal evolution time of
the open dynamical quantum system. This quantum speed limit time is
applicable to both mixed and pure initial states. We then apply this
result to the damped Jaynes-Cummings model and the Ohimc-like
dephasing model starting from a general time-evolution state. The
bound of this time-dependent state at any point in time can be
found. For the damped Jaynes-Cummings model, the corresponding bound
first decreases and then increases in the Markovian dynamics. While
in the non-Markovian regime, the speed limit time shows an
interesting periodic oscillatory behavior. For the case of
Ohimc-like dephasing model, this bound would be gradually trapped to
a fixed value. In addition, the roles of the relativistic effects on
the speed limit time for the observer in non-inertial frames are
discussed.

\end{abstract}

\maketitle

{\it{Introduction.}}---Quantum mechanics acting as a fundamental law
of nature imposes limit to the evolution speed of quantum systems.
The utility of these limits is shown in different scenarios,
including quantum communication \cite{Bekenstein}, the
identification of precision bounds in quantum metrology
\cite{Giovanetti}, the formulation of computational limits of
physical systems \cite{Lloyd}, as well as the development of quantum
optimal control algorithms \cite{Caneva}. The minimal time a system
needs to evolve from an initial state to its one orthogonal state is
defined as the quantum speed limit time (QSLT). The study of it has
been focused on both closed and open quantum systems. For closed
system with unitary evolution, a unified lower bound of QSLT is
obtained by Mandelstam-Tamm (MT) type bound and Margolus-Levitin
(ML) type bound
\cite{Mandelstam,Fleming,Anandan,Vaidman,Margolus,Levitin}. The
extensions of the MT and ML bounds to nonorthogonal states and to
driven systems have been investigated in Refs.
\cite{11,12,13,14,15,16}. The QSLT for nonunitary evolution of open
systems is also studied \cite{17,18,19}. It is shown that a unified
bound of QSLT including both MT and ML types for non-Markovian
dynamics can be formulated \cite{19}. However, while this unified
bound is applicable for a given driving time for the pure initial
states, it is not feasible for mixed initial states. As we all know
that decoherence and inaccurate operations are indispensable which
may result in mixed initial states.

In this Letter, we shall derive a QSLT for mixed initial states by
introducing relative purity as the distance measure, which can
characterize successfully the speed of evolution starting from an
arbitrary time-evolution state in the generic nonunitary open
dynamics. Let us consider the states of a driven system in the
damped Jaynes-Cummings model starting from a certain pure state
which corresponds to a special case of our result, one may observe
that the QSLT is equal to the driving time in the Markovian regime
\cite{19}. While by calculating the QSLT starting from the
time-evolution state at any point in time which is in general a
mixed state, it is interesting to find that the QSLT first begins to
decrease from the driving time and then gradually increases to this
driving time in the Markovian dynamics. So the speed of evolution in
the whole dynamical process exhibits an acceleration first and then
deceleration process. Additionally in the case of the non-Markovian
regime, the memory effect of the environment leads to a periodical
oscillatory behavior of the QSLT. We can also focus on the widely
used Ohmic-like reservoir spectra to investigate the QSLT for the
time-dependent states of the purely dephasing dynamics process. We
demonstrate that the QSLT will be reduced with the starting point in
time for Ohmic and sub-Ohmic dephasing model, that is to say, the
open system executes a speeded-up dynamics evolution process. While
for the super-Ohmic environments, due to the occurrence of coherence
trapping \cite{35}, we specifically point out that this QSLT would
be gradually trapped to a fixed value, and therefore leads to a
uniform evolution speed for the open system. We remark that the
findings of those phenomena rely on our general result of QSLT for
arbitrary initial states. Finally, we also investigate the influence
of the relativistic effect on the QSLT for the observer in
non-inertial frames in the above two quantum decoherence models.

{\it{Quantum speed limit time for mixed initial states.}}---In the
following, we shall consider a driven open quantum system and look
for the minimal time that is necessary for it evolve from a mixed
state $\rho_{\tau}$ to its final state $\rho_{\tau+\tau_{D}}$. Under
the general nonunitary quantum evolutions of open system, the final
state $\rho_{\tau+\tau_{D}}$ will be generally mixed. One general
choice of distance measure between two mixed states $\rho_{\tau}$
and $\rho_{\tau+\tau_{D}}$ is fidelity
$F(\rho_{\tau},\rho_{\tau+\tau_{D}})=tr[\sqrt{\sqrt{\rho_{\tau+\tau_{D}}}\rho_{\tau}\sqrt{\rho_{\tau+\tau_{D}}}}]$.
In this case of the initially mixed state $\rho_{\tau}$ should be
treated by purification in a sufficiently enlarged Hilbert space.
And the fidelity can be written
$F(\rho^{S}_{\tau},\rho^{S}_{\tau+\tau_{D}})=\max[|\langle\psi^{SE}_{\tau+\tau_{D}}|\varphi^{SE}_{\tau}\rangle|]$,
where the maximization is over all
$|\psi^{SE}_{\tau+\tau_{D}}\rangle$ $(|\varphi^{SE}_{\tau}\rangle)$
on a larger Hilbert space that are purifications of the mixed states
$\rho^{S}_{\tau+\tau_{D}}$ $(\rho^{S}_{\tau})$ on the smaller system
$S$. But performing the optimization over all possible purifications
is a challenging task that will be very hard to perform in the
general case.

Here we follow the relative purity as a distance measure to derive
lower bound on the QSLT for open quantum systems. The so-called
relative purity $f(\tau)$ between initial and final states of the
quantum system is defined as \cite{29}
$f(\tau+\tau_{D})=tr[\rho_{\tau+\tau_{D}}\rho_{\tau}]/tr(\rho^{2}_{\tau})$.
To evaluate the QSLT, let us now characterize the derivative of the
relative purity,
$\dot{f}(t)=tr[\rho_{\tau}\dot{\rho}_{t}]/tr(\rho^{2}_{\tau})$. The
rate of change of $f(t)$ will serve as the starting point for our
derivation to ML and MT type bounds on the minimal evolution time of
an initially mixed state $\rho_{\tau}$, using, respectively, the von
Neumann trace inequality and the Cauchy-Schwarz inequality.

By using the von Neumann trace inequality, we begin to provide a
derivation of ML type bound to arbitrary time-dependent nonunitary
equation of the form $\dot{\rho}_{t}=L_{t}(\rho_{t}).$ Let such a
map govern the evolution and consider
\begin{eqnarray}
\dot{f}(t)=\frac{tr[\rho_{\tau}L_{t}(\rho_{t})]}{tr(\rho^{2}_{\tau})}=\frac{tr[L_{t}(\rho_{t})\rho_{\tau}]}{tr(\rho^{2}_{\tau})}.\label{2}
\end{eqnarray}
Then, we introduce the von Neumann trace inequality for operators
which reads \cite{von Neumann,30},
$|tr({A_{1}A_{2}})|\leq\sum^{n}_{i=1}\sigma_{1,i}\sigma_{2,i}$,
where the above inequality holds for any complex $n{\times}n$
matrices $A_{1}$ and $A_{2}$ with descending singular values,
$\sigma_{1,1}\geq\ldots\geq\sigma_{1,n}$ and
$\sigma_{2,1}\geq\ldots\geq\sigma_{2,n}$. The singular values of an
operator $A$ are defined as the eigenvalues of $\sqrt{A^{\dag}A}$
\cite{30}. In the case of Hermitian operator, they are given by the
absolute value of the eigenvalues of $A$. We thus find $
|\dot{f}(t)|\leq\frac{1}{tr(\rho^{2}_{\tau})}\sum^{n}_{i=1}\sigma_{i}\varrho_{i}$,
with $\sigma_{i}$ are the singular values of $L_{t}(\rho_{t})$ and
$\varrho_{i}$ those of the initial mixed state $\rho_{\tau}$. Since
the singular values of $\rho_{\tau}$ satisfy $0<\varrho_{i}\leq1$,
the trace norm of $L_{t}(\rho_{t})$ would satisfy
$\|L_{t}(\rho_{t})\|_{tr}=\sum^{n}_{i=1}\sigma_{i}\geq\sum^{n}_{i=1}\sigma_{i}\varrho_{i}$,
so
\begin{eqnarray}
|\dot{f}(t)|\leq\frac{\sum^{n}_{i=1}\sigma_{i}\varrho_{i}}{tr(\rho^{2}_{\tau})}\leq\frac{\sum^{n}_{i=1}\sigma_{i}}{tr(\rho^{2}_{\tau})}.\label{4}
\end{eqnarray}
Integrating Eq. (\ref{4}) over time from $t=\tau$ to
$t=\tau+\tau_{D}$, we arrive at the ineqality
\begin{eqnarray}
\tau\geq\max\{\frac{1}{\overline{\sum^{n}_{i=1}\sigma_{i}\varrho_{i}}},\frac{1}{\overline{\sum^{n}_{i=1}\sigma_{i}}}\}|f(\tau+\tau_{D})-1|tr(\rho^{2}_{\tau}),\label{5}
\end{eqnarray}
where $\overline{X}=\tau_{D}^{-1}\int^{\tau+\tau_{D}}_{\tau}Xdt$.
For unitary processes, $\overline{\sum^{n}_{i=1}\sigma_{i}}$ is
equal to the time-averaged energy $E$, so the ML bound for closed
systems can be expressed
$\tau\geq\frac{1}{2\overline{\sum^{n}_{i=1}\sigma_{i}}}|f(\tau+\tau_{D})-1|tr(\rho^{2}_{\tau})$.

By noting the following inequality holds
$\sum^{n}_{i=1}\sigma_{i}\varrho_{i}\leq\sum^{n}_{i=1}\sigma_{i}$,
then
$\frac{1}{\overline{\sum^{n}_{i=1}\sigma_{i}\varrho_{i}}}\geq\frac{1}{\overline{\sum^{n}_{i=1}\sigma_{i}}}$.
So we can therefore simplify Eq. (\ref{5}) as
\begin{eqnarray}
\tau\geq{\frac{|f(\tau+\tau_{D})-1|tr(\rho^{2}_{\tau})}{\overline{\sum^{n}_{i=1}\sigma_{i}\varrho_{i}}}}.\label{6}
\end{eqnarray}
Regarding to general nonunitary open system dynamics, Eq. (\ref{6})
expresses a ML type bound on the speed of quantum evolution valid
for mixed initial states.

Next we want to derive a unified bound on the QSLT for the open
systems. According to Ref. \cite{18}, the rate of change of relative
purity can be bounded with the help of the Cauchy-Schwarz inequality
for operators,
$|tr({A_{1}A_{2}})|^{2}{\leq}tr(A^{\dag}_{1}A_{1})tr(A^{\dag}_{2}A_{2})$.
Then $
|\dot{f}(t)|\leq\frac{1}{tr(\rho^{2}_{\tau})}\sqrt{tr[L_{t}(\rho_{t})^{\dag}L_{t}(\rho_{t})]tr(\rho^{2}_{\tau})}$,
since $\rho_{\tau}$ is a mixed state, $tr(\rho^{2}_{\tau})<1$, and
we obtain
\begin{eqnarray}
|\dot{f}(t)|\leq\frac{\sqrt{tr[L_{t}(\rho_{t})^{\dag}L_{t}(\rho_{t})]}}{tr(\rho^{2}_{\tau})}=\frac{\|L_{t}(\rho_{t})\|}{tr(\rho^{2}_{\tau})},\label{8}
\end{eqnarray}
where
$\|L_{t}(\rho_{t})\|_{hs}=\sqrt{tr[L_{t}(\rho_{t})^{\dag}L_{t}(\rho_{t})]}=\sqrt{\sum^{n}_{i=1}\sigma^{2}_{i}}$
is the Hilbert-Schmidt norm. Integrating Eq. (\ref{8}) over time
leads to the following MT type bound for nonunitary dynamics
process,
\begin{eqnarray}
\tau\geq\frac{|f(\tau+\tau_{D})-1|tr(\rho^{2}_{\tau})}{\overline{\sqrt{\sum^{n}_{i=1}\sigma^{2}_{i}}}},\label{9}
\end{eqnarray}
where
$\overline{\sqrt{\sum^{n}_{i=1}\sigma^{2}_{i}}}=\tau_{D}^{-1}\int^{\tau+\tau_{D}}_{\tau}\sqrt{\sum^{n}_{i=1}\sigma^{2}_{i}}dt$
means the time-averaged variance of the energy.

Here, combining Eqs. (\ref{6}) and (\ref{9}), we obtain a unified
expression for the QSLT of arbitrary initially mixed states in open
systems, as following
\begin{eqnarray}
\tau_{QSL}=\max\{\frac{1}{\overline{\sum^{n}_{i=1}\sigma_{i}\varrho_{i}}},\frac{1}{\overline{\sqrt{\sum^{n}_{i=1}\sigma^{2}_{i}}}}\}|f(\tau+\tau_{D})-1|tr(\rho^{2}_{\tau}).\nonumber\\
\label{10}
\end{eqnarray}
Interestingly, for a pure initial state
$\rho_{\tau=0}=|\phi_{0}\rangle\langle\phi_{0}|$, the singular value
$\varrho_{i}=\delta_{i,1}$, then
$\sum^{n}_{i=1}\sigma_{i}\varrho_{i}=\sigma_{1}\leq\sqrt{\sum^{n}_{i=1}\sigma^{2}_{i}}$.
So expression (\ref{10}) thus reduces to the unified bound for the
QSLT has been given in Ref. \cite{19} based on fidelity, since
relative purity and fidelity are identical for pure initial states.
That is to say,  $\tau_{QSL}$ in expression (\ref{10}) can also be
defined as the minimal time a system needs to evolve from a pure
initial state to its final state.

{\it{The speed of evolution in the exactly solvable open system
dynamics.}}---In order to clear which bound on the speed limit time
$\tau_{QSL}$ can be attained and tight, we must compare
$\sum^{n}_{i=1}\sigma_{i}\varrho_{i}$ and
$\sqrt{\sum^{n}_{i=1}\sigma^{2}_{i}}$. In case
$\sum^{n}_{i=1}\sigma_{i}\varrho_{i}<\sqrt{\sum^{n}_{i=1}\sigma^{2}_{i}}$,
the ML type bound provides the tighter bound on the QSLT. The
unified expression (\ref{10}) of the speed limit time for mixed
initial states presented above is one of the results in this Letter.
Next we shall illustrate its use for the quantum evolution speed of
a qubit system in two decoherence channels. An generally mixed state
$\rho_{0}$ of a qubit can be written in terms of Pauli matrices,
whose coefficients define the so-called Bloch vector
$\rho_{0}=\frac{1}{2}(\mathbb{I}+v_{x}\sigma_{x}+v_{y}\sigma_{y}+v_{z}\sigma_{z})$,
where $\mathbb{I}$ is the identity operator of the qubit,
$\sigma_{k}$ $(k=x,y,z)$ is the Pauli operator, and
$v=(v_{x},v_{y},v_{z}){\in}B:=\{v\in\Re^{3}; \|v\|\leq1\}$.

We firstly consider the exactly solvable damped Jaynes-Cummings
model for a two-level system resonantly coupled to a leaky single
mode cavity. The environment is supposed to be initially in a vacuum
state. The nonunitary generator of the reduced dynamics of the
system is
$L_{t}(\rho_{t})=\gamma_{t}(\sigma_{-}\rho_{t}\sigma_{+}-\frac{1}{2}\sigma_{+}\sigma_{-}\rho_{t}-\frac{1}{2}\rho_{t}\sigma_{+}\sigma_{-})$,
where $\sigma_{\pm}=\sigma_{x}{\pm}i\sigma_{y}$ are the Pauli
operators and $\gamma_{t}$ the time-dependent decay rate. In the
case of only one excitation in the whole qubit-cavity system, the
environment can be described by an effective Lorentzian spectral
density of the form $
J(\omega)=\frac{1}{2\pi}\frac{\gamma_{0}\lambda}{(\omega_{0}-\omega)^{2}+\lambda^{2}}$,
where $\lambda$ is the width of the distribution, $\omega_{0}$
denots the frequency of the two-level system, and $\gamma_{0}$ the
coupling strength. Typically, weak-coupling regime
($\lambda>2\gamma_{0}$), where the behavior of the qubit-cavity
system is Markovian and irreversible decay occurs, and
strong-coupling regime ($\lambda<2\gamma_{0}$), where non-Markovian
dynamics occurs accompanied by an oscillatory reversible decay. The
time-dependent decay rate is then explicitly given by
$\gamma_{t}=\frac{2\gamma_{0}\lambda\sinh(dt/2)}{d\cosh(dt/2)+\lambda\sinh(dt/2)}$,
with $d=\sqrt{\lambda^{2}-2\gamma_{0}\lambda}$. The reduced density
opertor of the system at time $t$ reads
\begin{equation}
\rho_{t}=\frac{1}{2}\left(
       \begin{array}{cccc}
         2-(1-v_{z})p_{t} & (v_{x}-iv_{y})\sqrt{p_{t}} \\
         (v_{x}+iv_{y})\sqrt{p_{t}} & (1-v_{z})p_{t} \\
       \end{array}
     \right),\label{12}
\end{equation}
where $p_{t}=e^{-\int^{t}_{0}dt'\gamma_{t'}}$.
\begin{figure}[tbh]
\includegraphics*[width=6cm]{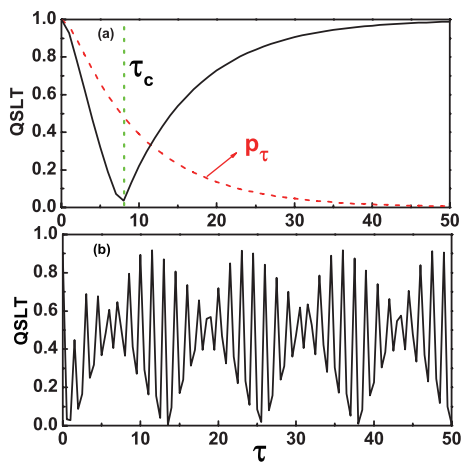}
\caption{ The QSLT for the damped Jaynes-Cummings model as a
function of the initial time parameter $\tau$ with $v_{z}=-1$,
$v_{x}=v_{y}=0$. (a) the Markovian regime, $\gamma_{0}=0.1\lambda$,
(b) the non-Markovian regime, $\gamma_{0}=10\lambda$. The red
(dashed) line represents the decay of the excited population
$p_{\tau}$. Parameters are $\omega_{0}=1$, $\lambda=1$ and
$\tau_{D}=1$.}
\end{figure}

 For the generally mixed state $\rho_{\tau}$ of a qubit,
$\varrho_{1}\sigma_{1}+\varrho_{2}\sigma_{2}$ is always less than
$\sqrt{\sigma^{2}_{1}+\sigma^{2}_{2}}$, so we reach the result that
the ML type bound on the QSLT is tight for the open system. The
unified expression (\ref{10}) proposed for the mixed initial states
in this Letter, can demonstrate the speed of the dynamics evolution
form an arbitrary time-dependent mixed state $\rho_{\tau}$ to
another $\rho_{\tau+\tau_{D}}$ by a driving time $\tau_{D}$. We
examine the whole dynamics process where the system starts in the
excited state, $v_{z}=-1$ and $v_{x}=v_{y}=0$. Figs. $1(a)$ and
$1(b)$ show the QSLT for a time-dependent mixed state $\rho_{\tau}$
as a function of $\tau$ in the Markovian and non-Markovian dynamics
process, respectively, in the case $\tau_{D}=1$. The QSLT can
initially reduce to a minimum, and gradually reach to the driving
time $\tau_{D}$ in the Markovian regime. While for the non-Markovian
regime, the speed limit time first decreases to a minimum in the
beginning of the evolution, then occurs a periodical oscillatory of
the time $\tau$. That is to say, in the Markovian regime, the
evolution of the qubit first exhibits a speeded-up process for
$\tau<\tau_{c}$ and then shows gradual deceleration process for
$\tau>\tau_{c}$. However, the speed of evolution for the qubit in
the non-Markovian dynamics process complies with an interesting
periodical oscillatory behavior.

The above behavior can be explained by evaluating the QSLT for the
qubit to evolve from $\rho_{\tau}$ to $\rho_{\tau+\tau_{D}}$,
\begin{equation}
\tau_{QSL}=\frac{|(p_{\tau}-p_{\tau+\tau_{D}})(1-2p_{\tau})|}{\frac{1}{\tau_{D}}\int^{\tau+\tau_{D}}_{\tau}|\dot{p}_{t}|dt}.\label{13}
\end{equation}
For the Markovian regime $\gamma_{t}=\gamma_{0}$, the value of
$|\dot{p}_{t}|$ can be given by $\gamma_{0}e^{-\gamma_{0}t}$, then
the speed limit time is simplified as
$\tau_{QSL}=\tau_{D}|1-2e^{-\gamma_{0}\tau}|$. So the appearance
seen in Fig. $1(a)$ depends only on the decay of the excited
population $p_{\tau}=e^{-\gamma_{0}\tau}$ for the time-dependent
state $\rho_{\tau}$, and the critical time
$\tau_{c}=\frac{1}{\gamma_{0}}\ln2$. Furthermore, the oscillatory
behavior shown by Fig. $1(b)$ in the the non-Markovian regime,
appears as a consequence of the oscillatory time dependence of the
decay rate $\gamma_{t}$.

In what follows, we consider a spin-boson-type Hamiltonian that
describes a pure dephasing type of interaction between a qubit and a
bosonic environment. It is worth stressing that this
qubit-plus-environment model admits an exact solution \cite{31,32}.
 There exists no correlations between the system and the environment at $t=0$; furthermore, the
environment is initially in its vacuum state at zero temperature.
The nonunitary generator of the reduced dynamics of the system is $
L_{t}(\rho_{t})=\gamma_{t}(\sigma_{z}\rho_{t}\sigma_{z}-\rho_{t})/2.$
By considering the bosonic environment operator is simply a sum of
linear couplings to the coordinates of a continuum of harmonic
oscillators described by a spectral function $J(\omega)$
\cite{33,34}, then
$\gamma_{t}=\int^{\infty}_{0}d{\omega}J(\omega)\coth(\frac{\hbar\omega}{2k_{B}T})\frac{1-\cos{\omega}t}{\omega^{2}}$.
Here, we suppose that the spectral density of the environmental
modes is Ohmic-like
$J(\omega)=\eta\frac{\omega^{s}}{\omega^{s-1}_{c}}e^{-\omega/\omega_{c}}$,
with $\omega_{c}$ being the cutoff frequency and $\eta$ a
dimensionless coupling constant. By changing the $s$-parameter one
goes from sub-Ohmic reservoirs ($s<1$) to Ohmic ($s=1$) and
super-Ohmic ($s>1$) reservoirs, respectively. For zero temperature,
$t>0$ and $s>0$, the dephasing rate can be obtained
$\gamma_{t}=\eta[1-\frac{\cos[(s-1)\arctan(\omega_{c}t)]\Gamma(s-1)}{(1+\omega^{2}_{c}t^{2})^{(s-1)/2}}]$,
where $\Gamma(s-1)$ is the Euler Gamma function. Taking the limit
$s\rightarrow1$ carefully, one also finds
$\gamma(t,s=1)=\eta\ln(1+\omega^{2}_{c}t^{2})$. The time evolution
of the reduced density matrix of the qubit satisfies
\begin{equation}
\rho_{t}=\frac{1}{2}\left(
       \begin{array}{cccc}
         1+v_{z} & (v_{x}-iv_{y})q_{t} \\
         (v_{x}+iv_{y})q_{t} & (1-v_{z}) \\
       \end{array}
     \right),\label{14}
\end{equation}
where $q_{t}=e^{-\gamma_{t}}$.

\begin{figure}[tbh]
\includegraphics*[width=6cm]{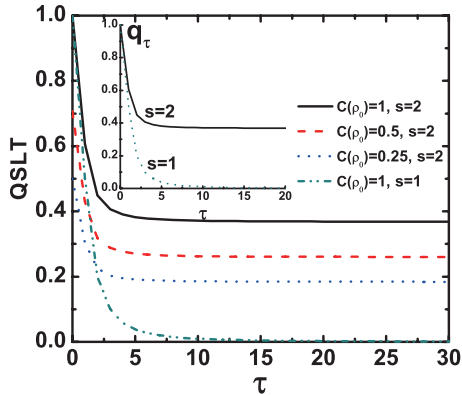}
\caption{The QSLT for the Ohmic-like dephasing model as a function
of the initial time parameter $\tau$ with the different coherence
$C(\rho_{0})$. The inset shows the decay rate $q_{\tau}$ of the
coherence for the mixed state $\rho_{\tau}$ with $C(\rho_{0})=1$.
Parameters are $\omega_{c}=1$, $\eta=1$ and $\tau_{D}=1$.}
\end{figure}

In this Ohmic-like dephasing model, the QSLT of a qubit can also be
given by ML type bound. In the dephasing evolution, by considering
an arbitrary mixed state $\rho_{\tau}$ to another
$\rho_{\tau+\tau_{D}}$ under a driving time $\tau_{D}$, the QSLT can
be calculated
\begin{equation}
\tau_{QSL}=\frac{C(\rho_{0})^{1/2}|q_{\tau}q_{\tau+\tau_{D}}-q^{2}_{\tau}|}{\frac{1}{\tau_{D}}\int^{\tau+\tau_{D}}_{\tau}|\dot{q}_{t}|dt}
,\label{15}
\end{equation}
with $C(\rho_{0})=v^{2}_{x}+v^{2}_{y}$ means the coherence of
$\rho_{0}$. With this, it is easy to show that $\tau_{QSL}$ is
independent of $v_{z}$, and not only relate to the dephasing rate of
the Ohmic-like environment but also to the coherence of the initial
state $\rho_{0}$ under a given driving time $\tau_{D}$. Fig. 2
presents the results of our analysis for $\tau_{QSL}$ in the
Ohmic-like dephasing process with different $C(\rho_{0})$. We
observe that, for the same driving time $\tau_{D}=1$, the lager
coherence of the initial state can decrease the speed of evolution
of a quantum system, and thus demand the longer QSLT. By choosing
the $s$-parameters satisfied Markovian regime \cite{PRARP}, the
speed limit time can be rewritten as
$\tau_{QSL}=\tau_{D}C(\rho_{0})^{1/2}q_{\tau}$. Hence, for a given
initial state $\rho_{0}$, the speed of evolution in the dynamics
process is determined by the decay rate $q_{\tau}$ of the coherence
for the mixed state $\rho_{\tau}$. Due to the specific form of the
spectral density for Ohmic-like dephasing model \cite{35}, in the
case of zero temperature, the qubit dephasing $q_{\tau}$ will
predict vanishing coherences in the long time limit for $s{\leq}1$.
On the other hand, for $s>1$ the qubit dephasing $q_{\tau}$ will
stop after a finite time, therefore leading to coherence trapping,
as shown by the inset of Fig. $2$. So the other notable observation
about Fig. $2$ is shown: The open system executes a speeded-up
dynamics evolution process in the Ohmic and sub-Ohmic dephasing
models. But for the super-Ohmic dephasing model the qubit firstly
exhibits a speeded-up dynamics process before a finite time, and
then complies with an uniform evolution speed after this finite
time.

{\it{Quantum speed limit time in non-inertial frames.}}---If the
observer for a quantum system in a uniformly accelerated frame with
acceleration $a$, the relativistic effect should be taken into
account \cite{20,21,22,25,26,27,28}. So here we shall investigate
the influence of the relativistic effect on the QSLT. Owing to the
relativistic effect, the coherence of the changed initial state
turns into ${\cos}^{2}r(v^{2}_{x}+v^{2}_{y})$, and becomes much less
than that of $C(\rho_{0})$. so the relativistic effect can increase
the speed of evolution of a quantum system in the purely Ohmic-like
dephasing channels. The parameter $r$ above is defined by
${\cos}r=(e^{-2\pi{\varpi}c/a}+1)^{-1/2}$, $c$ the speed of light in
the vacuum, and $\varpi$ the central frequency of the fermion wave
packet. But for the damped Jaynes-Cummings model, in spite of the
weaker coherence of the initial state brought by the relativistic
effect, the larger excited population
$1-\frac{(1+v_{z}){\cos}^{2}r}{2}$ in the changed initial state can
also be acquired. As well as the QSLT mainly depends on the
population of excited state under a given driving time in the
amplitude-damping channels \cite{36}, so the relativistic effect
would slow down the quantum evolution of the qubit in the damped
Jaynes-Cummings model, therefore leads to a smaller QSLT.

{\it{Conclusions.}}---We have derived a QSLT for arbitrary initial
states to characterize the speed of evolution for open systems. In
particular, considering the damped Jaynes-Cummings model, we have
obtained that the speed of evolution in the Markovian regime
exhibits an acceleration first and then deceleration process, and
shows a peculiar periodical oscillatory behavior in non-Markovian
regime. Moreover, in the case of the purely dephasing environments,
the QSLT would be gradually reduce to a fixed value for the
super-Ohmic dephasing model, and hence leads to a uniform evolution
speed for the open system. Our results may be of both theoretical
and experimental interests in exploring the speed of quantum
computation and information processing in the presence of noise.

This work was supported by ¡°973¡± program under grant No.
2010CB922904, the National Natural Science Foundation of China under
grant Nos. 11175248, 61178012, 11204156, 11304179 and 11247240.

\end{document}